\documentclass[allclo]{FBSart} 
\usepackage{amsfonts} 
\usepackage{amssymb} 
\usepackage{epsfig} 
\newcommand{\slsh}[1]{{\not \! #1}} 
 
\newcommand{\be}{\begin{equation}} 
\newcommand{\ee}{\end{equation}} 
\newcommand{\bea}{\begin{eqnarray}} 
\newcommand{\eea}{\end{eqnarray}} 
\newcommand{\M}{{\cal M}} 
\newcommand{\nn}{\nonumber}

\title{Truncated Schwinger-Dyson equations and gauge covariance in QED3}  
\author{A. Bashir\instnr{1,2,3}, A. Raya\instnr{1,2}} 
\instlist{Instituto de F{\'\i}sica y Matem\'aticas, 
Universidad Michoacana de San Nicol\'as de Hidalgo,  Morelia, Michoac\'an 58040, M\'exico. \and 
Instituto de Ciencias Nucleares, Universidad Nacional  
Aut\'onoma de M\'exico, Circuito Exterior,  
M\'exico, 
\and IPPP, Durham University, UK}  
 
\runningauthor{A. Bashir and A. Raya} 
\runningtitle{Truncated Schwinger-Dyson equations and gauge covariance in QED3} 
\begin{document} 
 
\maketitle 
 
\begin{abstract} 
We study the Landau-Khalatnikov-Fradkin transformations  
(LKFT) in momentum space for the dynamically generated mass function  
in QED3. Starting from the Landau gauge results in the rainbow approximation, we construct solutions in other covariant gauges. We confirm that the  
chiral condensate is gauge invariant as the structure of the LKFT predicts.  
We also check that the gauge dependence of the constituent fermion mass  
is considerably reduced as compared to the one obtained directly by solving  
SDE.  
\end{abstract} 
 
\section{Introduction} 
Gauge theories of fundamental interactions have been highly successful in collating experimental results in the perturbative regime. However, not all interesting phenomena can be understood in this approximation scheme. Confinement of quarks and gluons and the origin of hadronic masses are two examples. It is well known that on the distance scale of the order of hadron size, quarks behave as though their mass were 300 MeV, much larger than their vanishingly small current mass. In the context of covariant gauges, this effect may be attributed to an interplay of the behavior of gluon and ghost propagators in the infrared, \cite{Maris97}. These studies through Schwinger-Dyson equations (SDEs) have only been carried out in the Landau gauge. It is well known that if care is not taken in their truncation, solutions can be gauge dependent.  
 
Physically relevant strong dynamics is not restricted to QCD. There are also proposals of this kind to go beyond the standard model of particle physics such as (i) the technicolor models, (see\cite{Farhi} for an early review), (ii) top mode standard model and its modern variants (\cite{Hill} and references therein) including those with extra dimensions, {\it e.g.} \cite{Hashimoto1,Hashimoto2}. In view of these possibilities, it becomes all the more important that the corresponding studies of SDEs be independent of the ambiguities stemming from the approximations employed. 
 
Let alone the complicated battle ground of non-abelian theories, the ambiguities associated with the lack of gauge invariance in the non-perturbative truncations of SDEs in abelian theories such as QED are also a challenging problem. Among other issues, it has ultraviolet divergences. Most of the numerical investigations of SDEs make use of the more convenient  cut-off regularization method. However, such a cut-off can potentially introduce spurious gauge dependence \cite{Williams}, if employed naively. Consequently, one would hope that a neater testing ground for the validity of various truncation schemes could be provided by QED3, which is super-renormalizable. Nevertheless, a recent study,~\cite{FADM}, has revealed that the problem persists even in this seemingly simple scenario. The main motivation of this work is to address this problem of gauge non-invariance in QED3. 
 
In the absence of fermion loops, i.e., in the quenched approximation,  
the lattice version of QED3 has been shown to be confining~\cite{Gopfert1},  
making  its  connection to the strongly interacting QCD. However,  
if the vacuum polarization tensor is taken into account at order $\alpha$ 
with massless fermions circulating in the loop, the potential is no longer 
confining, see~\cite{quenched2}. Interestingly, for massive fermions, confinement is  
reinstated,~\cite{Burden1}. These results have received numerical confirmation 
in~\cite{Maris1}.  
 
In addition to being a toy model for SDE studies, QED3 is an attractive theory in its own right. In the field of superconductivity, see {e.g.}~\cite{supercon}, it has been used in the context of high $T_c$ superconductors.  These studies have received a boost from new experiments \cite{experiment}. QED3 has also found exciting applications in the recently explored unconventional quantum Hall effect in graphene \cite{gusexp}. Additionally, in the  realm of 
dynamical generation of fundamental fermion  masses, numerical findings   on the lattice, results obtained by employing  SDEs~\cite{differences} and alternative methods~\cite{mitra}, have yet to arrive at a final consensus, especially as regards the criticality of number of flavors, and continue to provide a popular battle ground.  
 
The problem of gauge invariance in QED3 can be traced back to not employing (or doing  so incorrectly) the gauge identities such as the Ward-Green-Takahashi identities (WGTI)~\cite{WGTI}, the  Nielsen identities~\cite{Nielsen} and the Landau-Khalatnikov-Fradkin transformations (LKFT)~\cite{LKF}. In this article, we address this issue  in the light of the LKFT. These transformations describe the specific manner in which Green functions transform under a 
variation of gauge. These  are non-perturbative in nature,  and are most readily defined in coordinate space. This was the basis for some earlier works~\cite{del}. Momentum space calculations are, however, more tedious~\cite{bashonly,bashLKF,delba,aitch}.  Due to their  non-perturbative nature, we expect LKFT not only to be satisfied at every order in perturbation theory, but also in phenomena which are realized non-perturbatively, such as dynamical  chiral symmetry breaking. Initial steps were taken in \cite{BRNPB1}   to apply these transformations directly to the dynamically generated mass function.  As a continuation of this effort, here we carry out an exact numerical exercise to 
study the behavior of the fermion propagator under these transformations in QED3. 
 
The traditional way to study gauge dependence is to make an {\em ansatz}  for the fermion-boson vertex, \cite{quenched2,BC,CP1,quenched3,BP1,BKP1,BKP2,BR1}, see \cite{nova} for a brief review. The fermion propagator is then calculated by solving the resulting SDE in different covariant  gauges.  This method is useful to test how consistently a truncation scheme performs in different gauges. However, this line of attack does not naturally guarantee that LKFT for the fermion propagator will be satisfied gauge by gauge. In order to ensure that these transformations are satisfied, it is obligatory to study the constraints imposed by them on the dynamically generated fermion propagator under a variation of gauge. One of the reasons it is important is because the said transformations guarantee that the condensate is a gauge-invariant quantity. Therefore they can provide a useful constraint on the truncation of SDEs. The way to do it is as follows. We start from the solution for the fermion propagator in the Landau gauge for a given truncation and  simply perform a LKFT to  find the result in any other gauge. The lesser the quantitative difference between the LKFT-generated fermion propagator and the one obtained by employing the truncation being tested in various gauges, the more reliable it is in terms of its gauge-covariant properties. 
  
We have organized this article as follows~: In Sect.~2  we review the generalities behind the truncation of SDE in QED in general, and QED3 in particular. Sect.~3 is devoted to the LKFT for the fermion propagator, and its behavior under gauge transformations as dictated by these transformations is reviewed in Sect.~4. Finally, we discuss our strategy and conclude in Sect.~5.

\begin{figure}[htbp] 
\begin{center} 
\includegraphics[width=0.8\columnwidth]{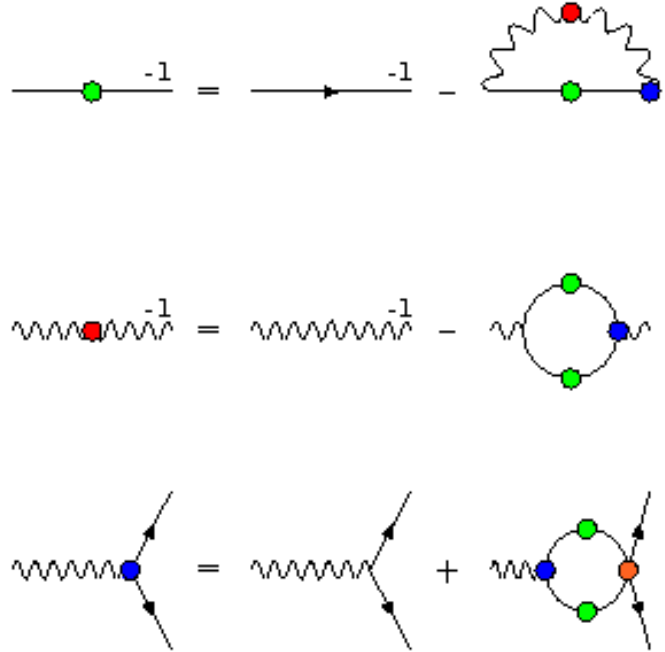} 
\caption{SDEs for the fermion propagator, photon propagator and fermion-boson vertex} 
\label{Fig:SDE-diagrams} 
\end{center} 
\end{figure} 
 
\section{Truncation of Schwinger-Dyson equations} 
 
We know that perturbation theory in QED does not generate fermion masses dynamically. Non-perturbative methods have to be employed  and SDEs are the natural tool for calculation in continuum. The first three equations of the infinite tower of SDEs in QED are depicted in Fig.~\ref{Fig:SDE-diagrams}.  
These correspond to the fermion propagator, the photon propagator and the fermion-boson vertex, respectively. The fermion propagator is coupled to the photon propagator and the fermion-boson vertex, which, through  their own SDE, are coupled to the rest of the infinite tower containing higher point Green functions.

\begin{figure}[htbp] 
\begin{center} 
\includegraphics[width=0.8\columnwidth,angle=-90]{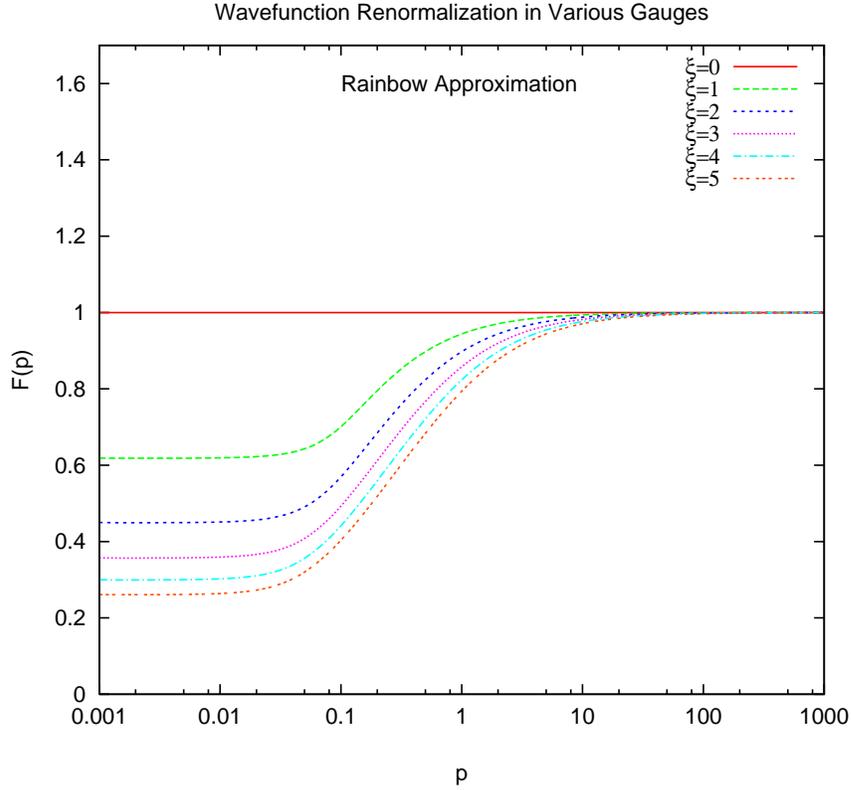} 
\caption{Wavefunction Renormalization in the Rainbow Approximation in various gauges. The scale of the graph is set by $e^2=1$} 
\label{Fig2} 
\end{center} 
\end{figure} 
 
\noindent 
The SDE for the fermion propagator in QED3 is
\bea 
S^{-1}(p;\xi)&=& S_0^{-1}(p)  
+e^2  
\int\frac{d^3k}{(2\pi)^3} 
\Gamma_\nu(k,p;\xi)S(k;\xi) \gamma_\mu\Delta_{\mu\nu}(q) \;, \; {} 
\label{SDEfp} 
\eea 
where $q=k-p$ and $e^2$ is the dimensionful coupling of QED3. In this expression $\Delta_{\mu\nu}(q)$ is  the photon propagator and $\Gamma_\nu(k,p;\xi)$  the full fermion-boson vertex. The most general form of the (Euclidean) fermion  propagator is  
\be 
S(p;\xi) = \frac{F(p;\xi)}{i {\not \! p}+{\cal M}(p;\xi)}\label{fpropmom} 
\ee 
where $F$ is referred to as the fermion wavefunction renormalization and $\cal M$ is the mass function.  
The photon propagator in its general form can be written as  
\bea 
\Delta_{\mu\nu}(q)&=& \frac{{\cal G}(q^2) }{q^2} 
\left( {\delta_{\mu\nu}}  - \frac{q_{\mu} q_{\nu}}{ q^2} \right)    + \xi  
\frac{q_\mu q_\nu}{q^4} 
 \equiv \Delta_{\mu\nu}^{T}(q) + \xi \frac{q_\mu q_\nu}{q^4}, 
\eea 
where $\cal G$ is the photon wavefunction renormalization, which is gauge invariant, and $\xi$ is the usual covariant gauge parameter.  
Expression~(\ref{SDEfp}) is a matrix equation which can be converted into system of coupled non-linear scalar integral equations for $\cal M$ and $F$ after multiplying it, respectively, by 1 and $\slsh{p}$ and taking trace.   A favorite starting point in the quenched version of QED3, which consists in neglecting fermion loops (${\cal G}=1$), is to choose a suitable \emph{ansatz} for the fermion-boson vertex, in such a way that the SDE for the fermion propagator can be solved  for the unknowns $\M$ and $F$.  
Possibly the simplest choice for the vertex, which allows us to understand the general features of dynamical chiral symmetry breaking, consists in replacing  the fermion-boson vertex by its bare counterpart~\cite{quenched2,BHR}. This is the so-called rainbow approximation.

\begin{figure}[htbp] 
\begin{center} 
\includegraphics[width=0.8\columnwidth,angle=-90]{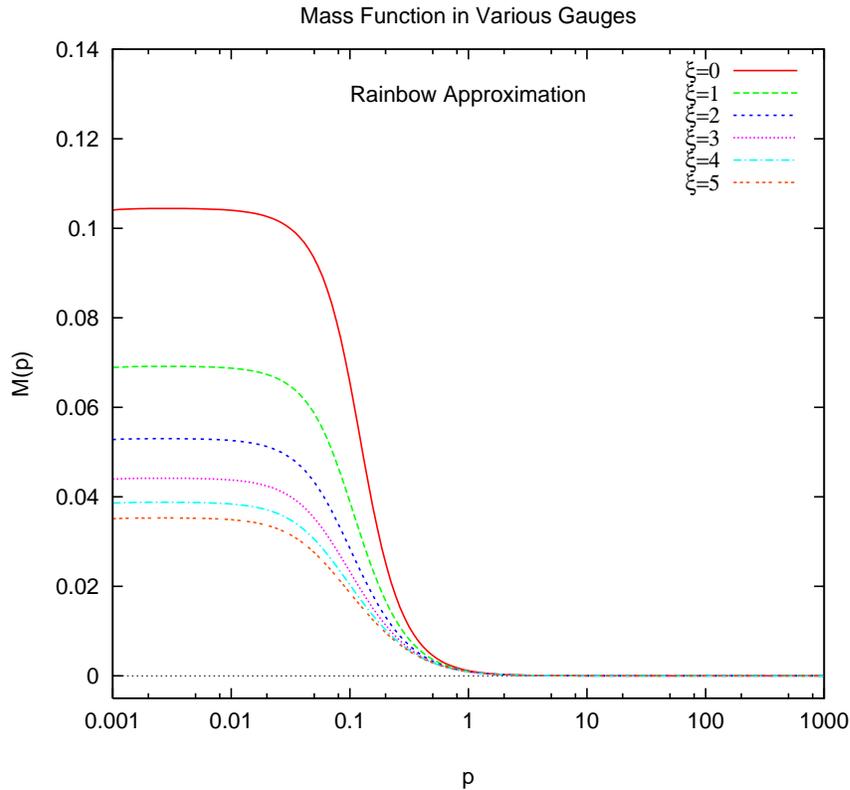} 
\caption{Mass Function in the Rainbow Approximation in various gauges. The scale of the graph is set by $e^2=1$} 
\label{Fig3} 
\end{center} 
\end{figure} 
 
\begin{figure}[htbp] 
\begin{center} 
\includegraphics[width=0.8\columnwidth,angle=-90]{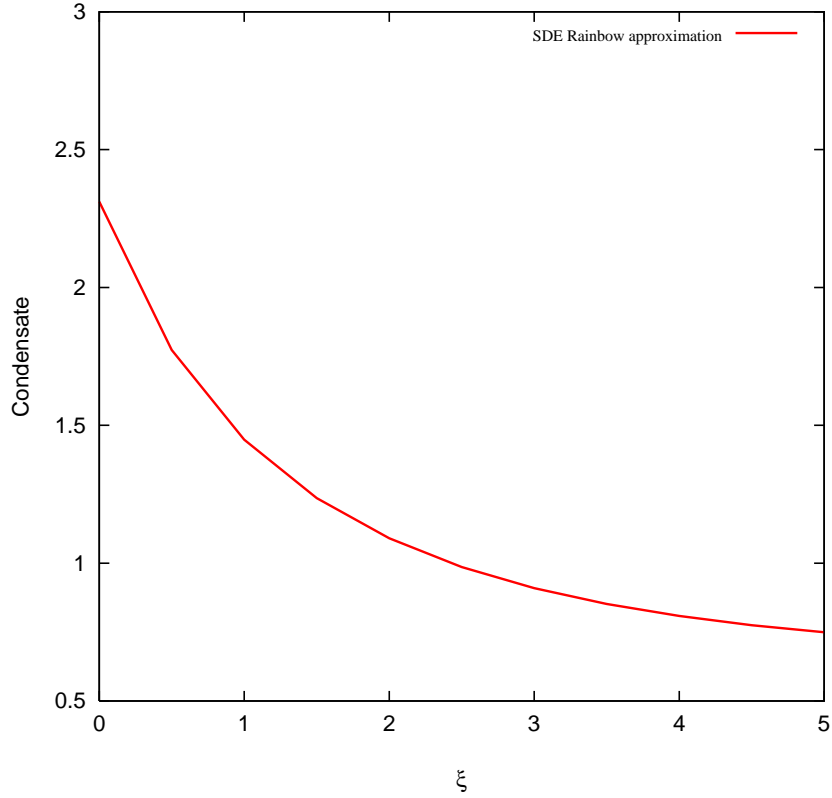} 
\caption{Chiral condensate in the Rainbow approximation. The scale of the graph is given in units of $10^{-3}e^4=1$} 
\label{fcond} 
\end{center} 
\end{figure}

\noindent 
To start with, the bare fermion propagator is considered massless. In this case the unknown functions defining the fermion propagator are found through the following system of equations, 
\begin{eqnarray} 
\frac{1}{F(p;\xi)}&=& 1-\frac{\alpha\xi}{4\pi}\int_0^\infty dk 
\frac{k^2F(k;\xi)}{k^2+\M^2(k;\xi)}
\left[1-\frac{k^2+p^2}{2kp}\ln{\left| 
\frac{k+p}{k-p}\right|} \right],\nn\\ 
\frac{\M(p;\xi)}{F(p;\xi)}&=& \frac{\alpha(\xi+2)}{\pi p} \int_0^\infty dk 
\frac{k F(k;\xi) \M(k;\xi)}{k^2+\M^2(k;\xi)}
\ln{\left| \frac{k+p}{k-p}\right|}, 
\label{SDErainbowintegrated} 
\end{eqnarray} 
where $\alpha=e^2/4\pi$ as usual. In Figs.~\ref{Fig2} and~\ref{Fig3} we show the functions $F$ and $\cal M$ as a function of momentum for various values of the gauge parameter.  
$\M(p;\xi)$ is roughly a constant for low momentum and falls as $1/p^2$ as $p\to\infty$ \cite{BHR}, for every value of $\xi$, whereas $F(p;\xi)$ is a constant different from unity in the infrared  in all but the Landau gauge, and approaches  1 at large $p$ for all values of the gauge parameter.   
The non-trivial profile of the mass function is a result of chiral symmetry being dynamically broken. Observe that the curves corresponding to ${\cal M}(p;\xi)$ never cross each other for different values of $\xi$. The resulting physical observables such as the chiral condensate $\langle\bar\psi\psi\rangle = - {\rm Tr} S(x=0;\xi),$ are gauge-dependent quantities, as can be seen in Fig~\ref{fcond}.  
One can expect the source of such gauge dependence the fact that the bare vertex violates the WGTI, 
\be 
iq_{\mu} \Gamma_\mu(k,p;\xi) = S^{-1}(k;\xi) - S^{-1}(p;\xi)\;. 
\ee 

In order to recover the reliability of a truncation to the SDE, one models the form of the three-point vertex, requiring, for example, the validity of the WGTI. Formally, in light of this identity, the vertex can be decomposed into two parts,
\begin{equation} 
\Gamma_{\mu}(k,p;\xi) = \Gamma^L_{\mu}(k,p;\xi) + \Gamma_{\mu}^T(k,p;\xi)\;, 
\end{equation}  
such that $q_{\mu} \Gamma_{\mu}^T(k,p;\xi)=0$. $\Gamma_{\mu}^T(k,p;\xi)$ remains undetermined by the WGTI and is called  
the transverse part. $\Gamma^L_{\mu}(k,p;\xi)$ is the  longitudinal part. Although it is not uniquely fixed by the WGTI, a popular choice  is the so-called Ball-Chiu vertex, Eq.~(3.2) in~\cite{BC}. An {\em ansatz} can be made for the transverse part of the vertex guided by perturbation theory, by the LKFT transformations, freedom of kinematic singularities and its correct transformation properties under $C$, $P$ and $T$. A few examples in this connection are the vertices proposed by Bashir, Curtis and Pennington,~\cite{CP1,BP1}. Yet the question remains, what should one expect to be the behavior of the fermion propagator in different gauges that ensures the gauge independence of physical observables associated with the phenomenon of DCSB?  Whereas WGTI ensures the gauge invariance of the pole mass (in Minkowski space), LKFT preserve the gauge independence of the chiral condensate. Therefore, in addition to the constraints of the WGTI, the ones of LKFT can also guide us toward improved truncations of SDEs. This is what we concretize in the next section. 
 
\section{LKFT and the Fermion Propagator} 
 
We start by writing the Euclidean space fermion propagator in both momentum and  coordinate spaces in their most general forms~: 
\bea 
S(p;\xi) \equiv 
\frac{F(p;\xi)}{i {\not \! p}+{\cal M}(p;\xi)} \;,  \qquad 
S(x;\xi) \equiv {\not \! x}X(x;\xi)+ Y(x;\xi) \;. 
\label{fpropcoord}  
\eea 
Expressions in Eq.~(\ref{fpropcoord}) are related through a $d-$dimensional Fourier transformation. Assuming we know the fermion propagator in Landau gauge in momentum space, $S(p;0)$,  the LKFT relating the coordinate space fermion propagator in the Landau gauge to the one in an arbitrary covariant gauge reads 
\begin{eqnarray} 
S(x;\xi) &=& S(x;0) \; \exp{\left[-i \left[ \Delta_d(0) - \Delta_d(x)  \right]\right]} \;, 
\end{eqnarray} 
where  
\begin{eqnarray} 
	\Delta_d(x) &=& -i \xi e^2 \; \int \frac{d^dp}{(2 \pi)^d} \; \frac{\exp{\left[-i p \cdot x\right]}}{p^4} \;. 
\end{eqnarray} 
This transformation leaves SDEs and WGTI functionally invariant in different gauges, and thus imposes strong constraints on the gauge covariance of the dynamically generated fermion propagator. It is trivial to see that  corresponding LKFT  ensures the gauge invariance of the chiral condensate: 
\begin{eqnarray}  
\langle\bar\psi\psi\rangle_\xi & =& - {\rm Tr} S(x=0;\xi)\nn\\ 
&=&- {\rm Tr} S(x=0;0)\; \exp{\left[-i \left[ \Delta_d(0) - \Delta_d(x=0)  \right]\right]}\nn\\ 
&=&-{\rm Tr} S(x=0;0) = \langle\bar\psi\psi\rangle_0\;. 
\end{eqnarray} 
Thus, the LKFT in SDEs studies should play an important role to address the issue of gauge invariance. Below we exemplify this considering QED3.

\begin{figure}[htbp] 
\begin{center} 
\includegraphics[width=0.8\columnwidth,angle=-90]{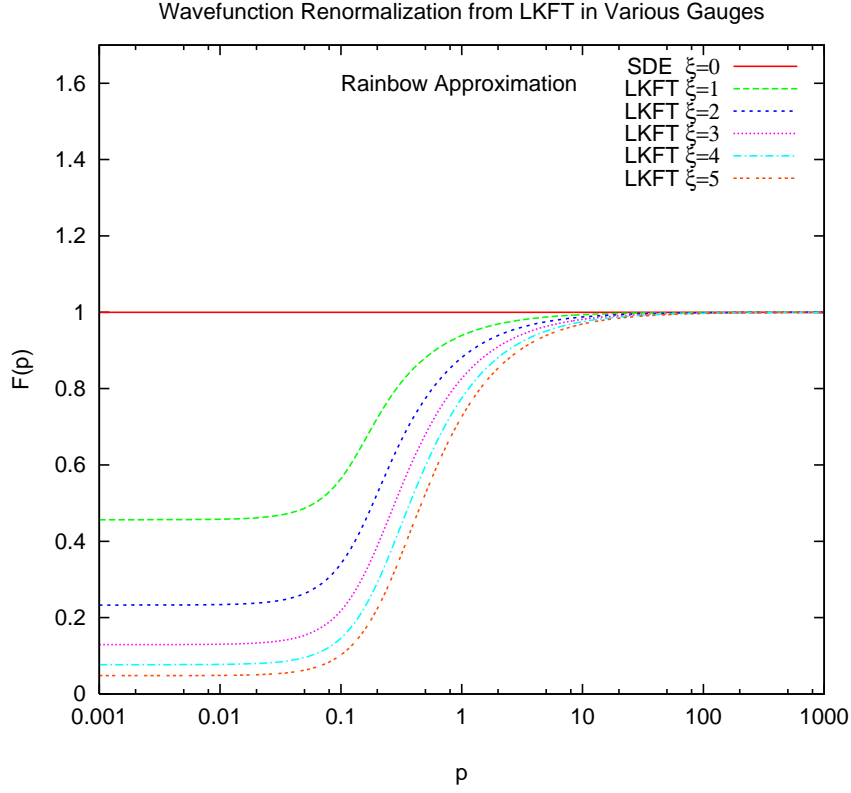} 
\caption{Wavefunction Renormalization in the Rainbow Approximation in various gauges from LKFT. The scale of the graph is set by $e^2=1$} 
\label{lkf-F} 
\end{center} 
\end{figure} 
 
The dynamically generated fermion propagator is usually known in momentum space. We can obtain a momentum-space representation for the LKFT in the following manner. Assuming we know  $S(p;0)$, we have the corresponding expressions for the propagator in coordinate space given by 
\bea 
X(x;0)&=&-\frac{i}{x^2}\int \frac{d^3k}{(2\pi)^3}  
\frac{F(k;0)}{k^2+{\cal M}^2(k;0)} \ \exp{\left[-ik\cdot x\right]} k\cdot x \;,\nn\\ 
Y(x;0)&=& - \int \frac{d^3k}{(2\pi)^3}    
\frac{F(k;0)\ {\cal M}(k;0)}{k^2+{\cal M}^2(k;0)} \ \exp{\left[-ik\cdot x\right]}\;. 
\eea 
Performing the angular integrations, these expressions get simplified as follows, 
\bea 
X(x;0)&=& \frac{1}{2\pi^2x^2}\int_0^\infty dk \frac{k^2F(k;0)}{k^2+{\cal M}^2(k;0)}
\left[\cos{kx} -\frac{\sin{kx}}{kx} \right]\nn\\ 
Y(x;0)&=&- \frac{1}{2\pi^2}\int_0^\infty \! dk k^2 
\frac{ F(k;0)\ {\cal M}(k;0)}{k^2+{\cal M}^2(k;0)} 
\left[ \frac{\sin{kx}}{kx}\right].\label{XY0} 
\eea 
It is straightforward to check that for  $d=3$, $\Delta_3(0) - \Delta_3(x) = -i \alpha \xi /2$. Hence, the LKFT transformation rule in QED3 simplifies to 
\begin{eqnarray} 
S(x;\xi)&=&S(x;0)\;e^{-a x} \;, 
\end{eqnarray} 
with $a=\alpha \xi/2$. 
After applying LKFT to the expressions in Eq.~(\ref{XY0}) and Fourier transforming the results back to momentum space, we get 
\bea 
\frac{F(p;\xi)}{p^2+{\cal M}^2(p;\xi)}&=& \frac{i}{p}\int d^3x \ \exp{\left[ip\cdot x\right]} X(x;\xi)\;,\nn\\ 
\frac{ F(p;\xi){\cal M}(p;\xi)}{p^2+{\cal M}^2(p;\xi)}&=& -\int d^3x \ \exp{\left[ip\cdot x\right]} Y(x;\xi)\;. 
\eea

\begin{figure}[htbp] 
\begin{center} 
\includegraphics[width=0.8\columnwidth,angle=-90]{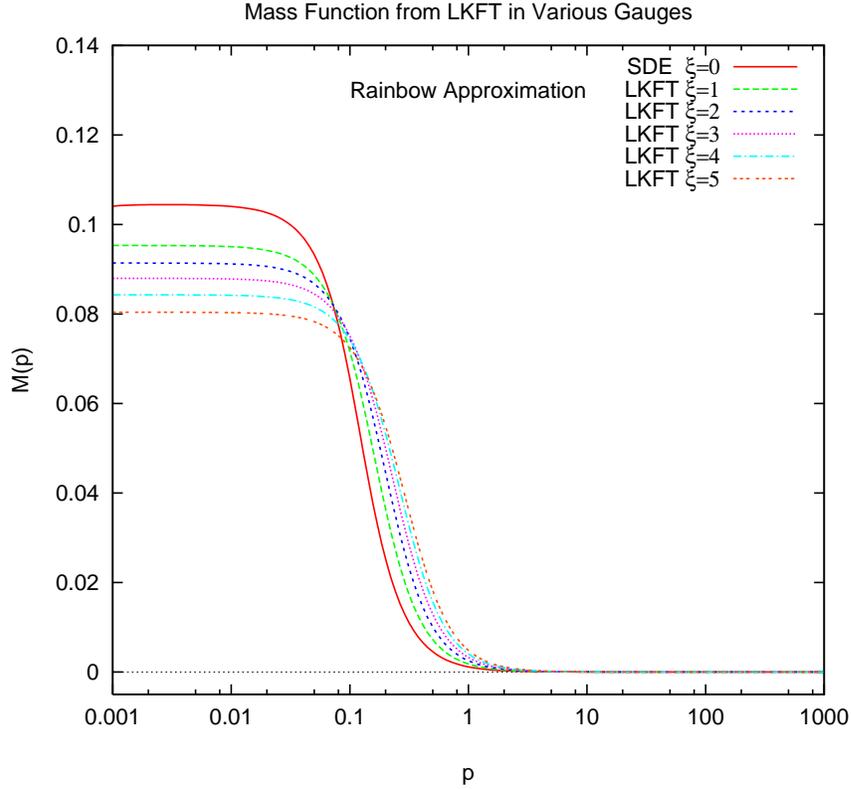} 
\caption{Mass Function in the Rainbow Approximation  in various gauges from LKFT. The scale of the graph is set by $e^2=1$} 
\label{lkf-M} 
\end{center} 
\end{figure} 

\noindent 
Angular integrations lead to  the final momentum space representation of the LKFT for the fermion propagator (provided $a>0$)~: 
\bea 
\frac{F(p;\xi)}{p^2+\M^2(p;\xi)} &=&  
\frac{a}{\pi p^2}\int_0^\infty dk\, k^2 \frac{F(k;0)}{k^2+\M^2(k;0)} 
\left[\frac{1}{\lambda^ -}+\frac{1}{\lambda^+}+\frac{1}{2kp} 
\ln{ \left| {\frac{\lambda^-}{\lambda^+}} \right| } 
\right], \nn \\ \nn \\ 
\frac{F(p;\xi)\ \M (p;\xi)}{p^2+\M^2(p;\xi)} &=&  
\frac{a}{\pi p}\int_0^\infty dk\, k \frac{F(k;0)\ \M (k;0)}{k^2+\M^2(k;0)} 
\left[\frac{1}{\lambda^-}-\frac{1}{\lambda^+}\right]\;, \label{gist} 
\eea 
where $\lambda^\pm = a^2 + (k\pm p)^2$.   
Thus knowledge of $S(p;0)$ is the input required to obtain the same in an arbitrary covariant gauge. One can further confirm that this representation preserves the gauge invariance of the chiral condensate upon integration the second expression over $\int d^3p/(2\pi)^3$. The left-hand side corresponds to the integrand of the condensate in momentum space.

\section{Behavior of the fermion propagator} 
 
We shall now review the behavior of the fermion propagator in different gauges as dictated by the corresponding LKFT. 
Starting with the chirally asymmetric solution of the fermion propagator in the Landau gauge,  we apply the LKFT to obtain the off-shell fermion propagator in an arbitrary covariant gauge.

\begin{figure}[htbp] 
\begin{center} 
\includegraphics[width=0.8\columnwidth,angle=-90]{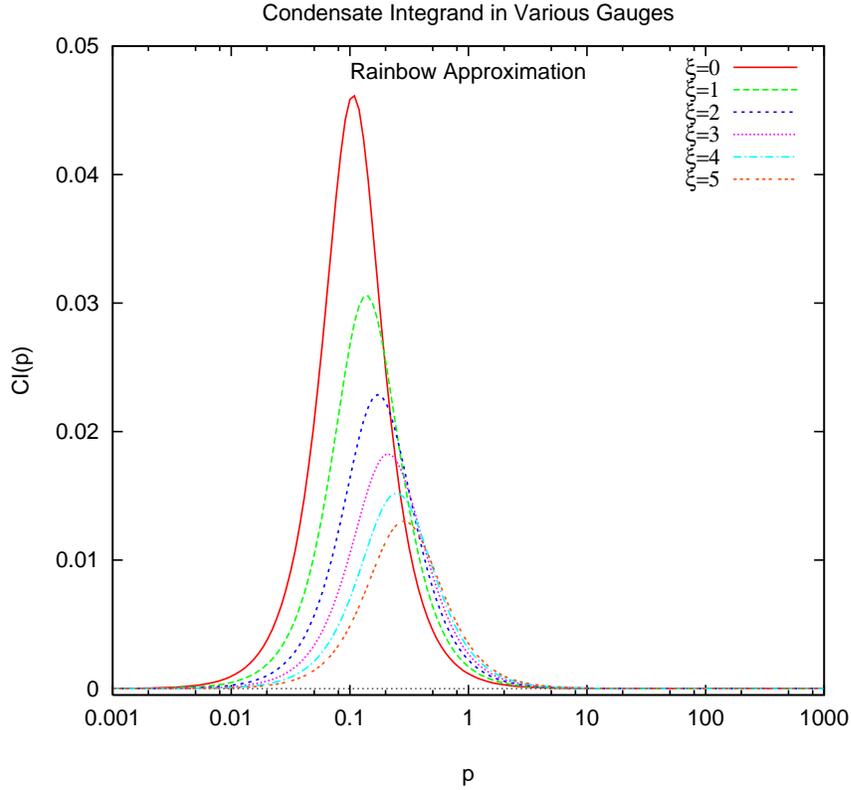} 
\caption{Condensate integrand in the Rainbow Approximation  in various gauges from LKFT. The scale of the graph is set by $e^2=1$} 
\label{CI} 
\end{center} 
\end{figure} 
 
\begin{figure}[htbp] 
\begin{center} 
\includegraphics[width=0.8\columnwidth,angle=-90]{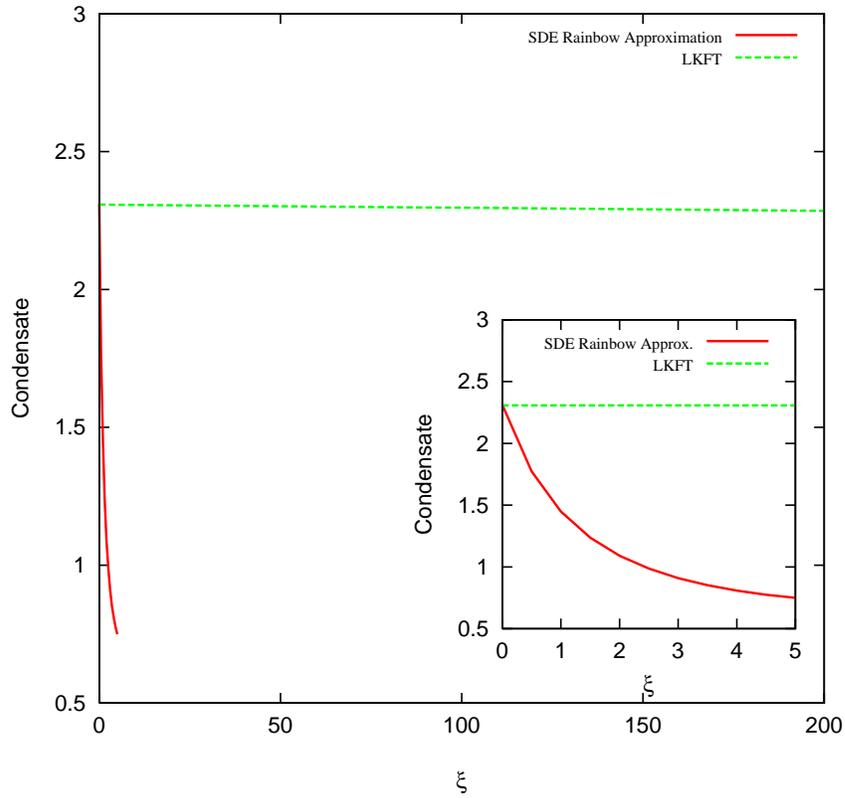} 
\caption{Gauge dependence of the condensate in the Rainbow Approximation from SDE and LKFT. The scale of the graph is set by $e^2=1$} 
\label{conden} 
\end{center} 
\end{figure}

\noindent 
The behavior of the fermion propagator is depicted in Figs~\ref{lkf-F} and~\ref{lkf-M}. Observe that  the wavefunction renormalization in qualitatively the same both for the SDE solution in different gauges and the LKF-transformed solution. For the case of the mass function, the asymptotic behavior remains qualitatively the same, as the analytical treatment in Ref.~\cite{BRNPB1} confirms, but the key observation is something that might have been anticipated but has not before been demonstrated; namely, a rearrangement in the mass function in different gauges in such a fashion as to render the chiral condensate gauge invariant.  
This is illustrated in Fig.~\ref{CI}, where we depicted the condensate integrand  
\begin{equation} 
CI(p) = \frac{p^2 F(p;\xi)\ \M(p;\xi) }{p^2+\M^2(p;\xi)} 
\end{equation} 
as a function of momentum for various values of the gauge parameter. A simple numerical exercise reveals that the area under each curve is equal, rendering the condensate gauge invariant. This can explicitly be seen in Fig.~\ref{conden}, where the flat line corresponds to the gauge (in)dependence of the condensate calculated from the LKF transformed solutions as compared with the SDE results.

Constraints on the behavior of the fermion propagator arising from LKFT  
have implications on other physical observables, like the constituent mass  
in Euclidean space defined by $m = \M(p = m)$. A comparison from the results obtained  
from the rainbow truncation of SDE [31] and those coming from LKFT is depicted in Fig. 9. Gauge dependence of the constituent mass arising from LKFT  
has been reduced considerably, as compared with the sharp gauge dependence  
of the SDE result.  
 
\begin{figure}[htbp] 
\begin{center} 
\includegraphics[width=0.8\columnwidth,angle=-90]{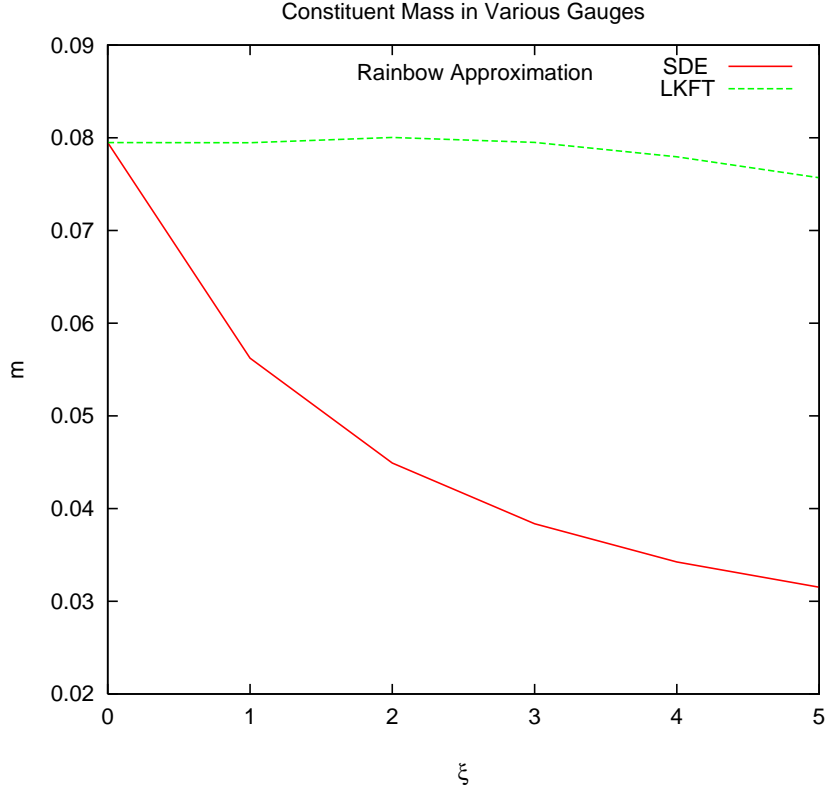} 
\caption{Constituent mass in the Rainbow Approximation  in various gauges from LKFT. The scale of the graph is set by $e^2=1$} 
\label{Cmass} 
\end{center} 
\end{figure} 
 
  It has been shown that the fermion propagator has complex mass like singularities implying confinement~\cite{Maris1}. Through the evaluation of the 
Euclidean-time Schwinger function, one can determine whether or not the fermion propagator corresponds to a physically observable state or  
not~\cite{HRM,HRW}. It is defined as
\begin{eqnarray} 
	 \Delta(t;\xi) &=& \int d^2 x \; \int \frac{d^3p}{(2 \pi)^3} \; \exp{\left[i(p_0 t+ \mathbf{p} \cdot \mathbf{x})\right]} \; \sigma(p^2;\xi)  \;, 
\end{eqnarray} 
where 
\begin{eqnarray} 
    	 \sigma(p^2;\xi) &\equiv& \frac{F(p;\xi) \, \M(p;\xi)}{p^2+\M^2(p;\xi)} \;. 
\end{eqnarray} 
It can be shown that if there is a stable asymptotic state of mass $m$ associated with this propagator, then  
\begin{eqnarray*} 
	    \Delta(t;\xi) \simeq {\rm e}^{-mt} 
\end{eqnarray*} 
for large Euclidean $t$. This readily implies 
\begin{eqnarray} 
	- \lim_{t \rightarrow \infty} \frac{d}{dt} \; {\rm ln} [\Delta(t,\xi)] = m \;. 
\end{eqnarray} 
On the other hand, two complex conjugate mass like singularities with complex masses $\mu = a \pm i b$ lead to an oscillating behavior such as  
\begin{eqnarray} 
	     \Delta(t;\xi) \simeq {\rm e}^{-at} \; {\rm cos} (bt + \delta)  
\end{eqnarray} 
for large $t$. If we want to associate physical interpretation to these masses, there gauge independence needs to be 
established. As the above-mentioned method is not very accurate, associating exact numbers to $a$ and $b$ in various  
gauges might not reflect the gauge (in)dependence of these quantities. However, we present a qualitative idea through 
Fig. 10. 
 
\begin{figure}[htbp] 
\begin{center} 
\includegraphics[width=0.8\columnwidth,angle=-90]{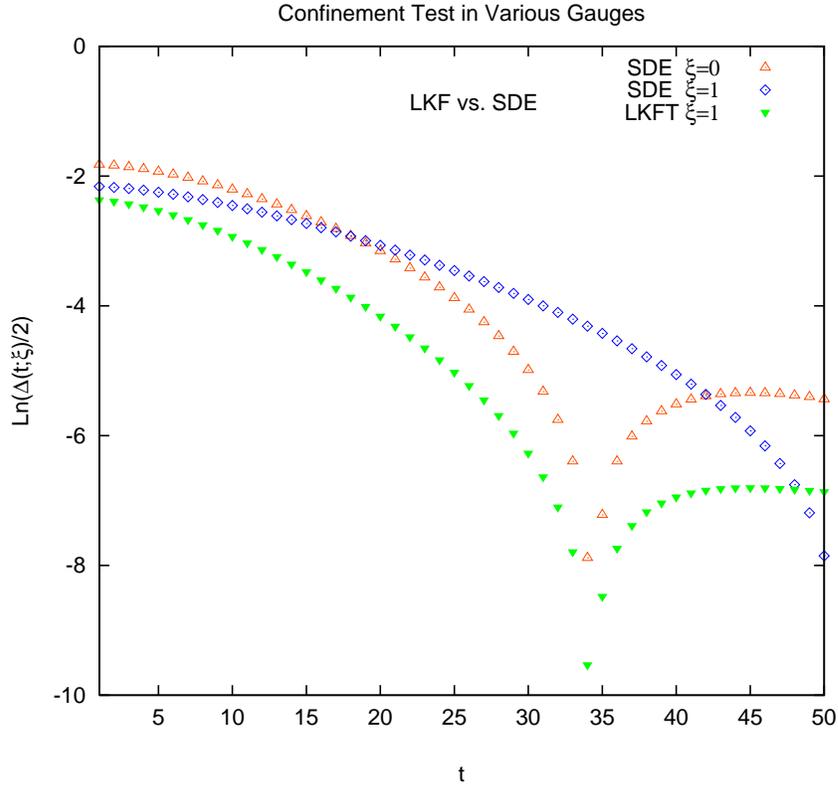} 
\caption{${\rm ln}[\Delta(t;\xi)/2]$ for $\xi=0,1$ through SDE and LKFT generated mass function} 
\end{center} 
\end{figure} 
 
\noindent 
We see that the LKFT-generated ${\rm ln}[\Delta(t;1)/2]$ is qualitatively much closer to ${\rm ln}[\Delta(t;0)]$ than the 
SDE-generated ${\rm ln}[\Delta(t;1)]/2$. Exact numbers require solving SDE for complex momenta.

\section{Discussion and Conclusions} 
 
\begin{figure}[htbp] 
\begin{center} 
\includegraphics[width=0.8\columnwidth,angle=-90]{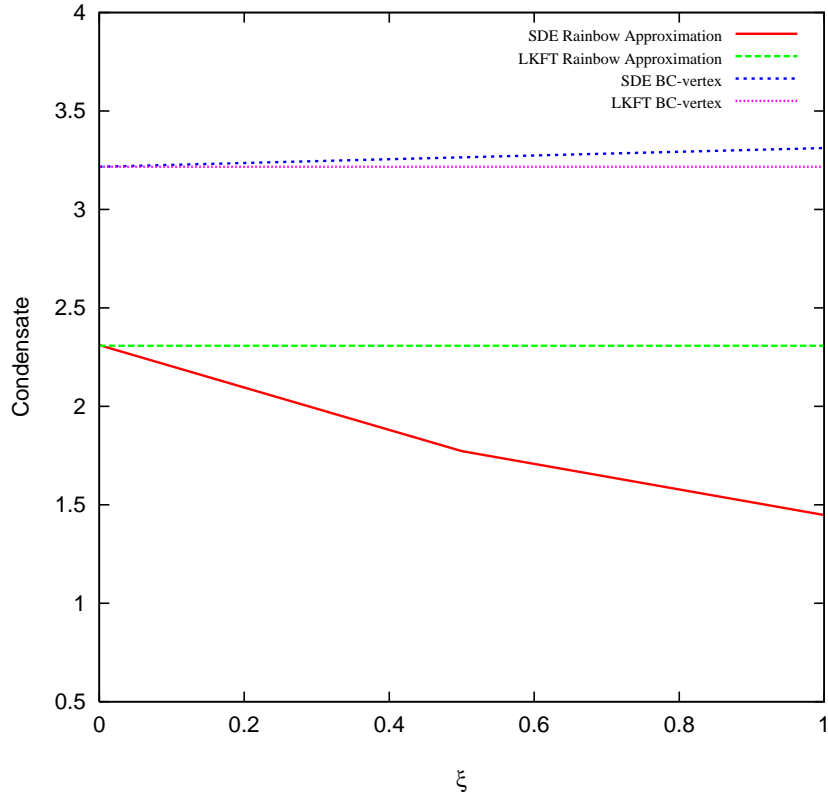} 
\caption{Comparison of the gauge dependence of the condensate for the Rainbow Approximation and the Ball-Chiu vertex from SDE and LKFT. The scale of the graph is set by $e^2=1$} 
\label{ball} 
\end{center} 
\end{figure}

To conclude, just as WGTI provides useful constraints on the truncations of SDEs, so do the LKF transformations. It is constructive to compare the solutions of  the SDE in different gauges against those generated by the LKFT, \cite{BRNPB1}. If the off-shell propagators obtained through these two methods differ drastically from each other, the truncation adopted does not conform to the gauge covariance properties of the fermion propagator. For example, in Fig.~\ref{ball} we compare the gauge dependence of the condensate for two different truncation schemes with our strategy. On the one hand we show the rainbow approximation, where results from SDE and LKFT show a clear difference. On the other hand, we show the same comparison for the Ball-Chiu vertex~\cite{BC}. This proposal fulfills the WGTI, hence rendering a less pronounced gauge dependence for the condensate, as can be seen from the small difference between SDE and LKFT results.  
This procedure detailed in the paper is also valid when one goes beyond the quenched truncation and tries to calculate the vacuum polarization. The same vertex must appear in both the 
fermion and photon gap equations. The bare vertex is inadequate because the vacuum polarization tensor must be transverse. However, once a vertex is constructed to ensure this, LKFT 
will provide the vertex required in another gauge. 
Therefore, more definitive conclusions upon the most appropriate vertex \emph{ansatz} can be drawn after a detailed study of the LKFT for the vertex function itself~\cite{BRP}. This work is in progress.  A natural extension of this work is to study the gauge covariance of the quark propagator through the generalized LKFT  of QCD.

\begin{acknowledge} 
We acknowledge fruitful discussions with M.R. Pennington at all stages of this work. We thank V.P. Gusynin, C.S. Fischer, R. Williams  and C.D. Roberts for valuable discussions. A.R. is grateful for the hospitality of the IPPP at Durham University during his visit.   AB wishes to acknowledge a short term visitor grant by a joint scheme  of The Royal Society, U.K, and The Mexican Academy of Sciences, Mexico. Support has also been provided by CIC (grants 4.10 and 4.22) and CONACyT  (grant 46614-I). 
\end{acknowledge}

\end{document}